\documentclass[preprint,12pt]{elsarticle}

 \usepackage{graphics}
 \usepackage{graphicx}

\usepackage{amssymb}
 \usepackage{amsthm}
\usepackage{hyperref}
\usepackage{color}
\usepackage{xcolor}
\usepackage{graphicx}
\usepackage{multirow}
\usepackage{amsmath,amssymb,amsthm,amsxtra,overpic,bbm,bm,float
,epsfig}

\newcounter{bla}

\journal{Computer Physics Communications}

\begin{document}

\begin{frontmatter}



\title{Flavour Symmetry Embedded - GLoBES (FaSE-GLoBES)}


\author[a]{Jian Tang}
\author[a]{TseChun Wang\corref{author}}

\address[a]{School of Physics, Sun Yat-Sen University, Guangzhou 510275, China}

\begin{abstract}
Neutrino models based on flavour symmetries provide the natural way to explain the origin of tiny neutrino masses. At the dawn of precision measurements of neutrino mixing parameters, neutrino mass models can be constrained and examined by on-going and up-coming neutrino experiments. We present a supplemental tool \textbf{F}l\textbf{a}vour \textbf{S}ymmetry \textbf{E}mbedded (\textbf{FaSE}) for \textbf{G}eneral \textbf{Lo}ng \textbf{B}aseline \textbf{E}xperiment \textbf{S}imulator (\textbf{GLoBES}), and it is available via the link \url{https://github.com/tcwphy/FASE_GLoBES}. It can translate the neutrino mass model parameters to standard neutrino oscillation parameters and offer prior functions in a user-friendly way. We demonstrate the robustness of \textbf{FaSE-GLoBE} with four examples on how the model parameters can be constrained and even whether the model is excluded by an experiment or not. We wish that this toolkit will facilitate the study of new neutrino mass models in an effecient and effective manner.
\end{abstract}


\end{frontmatter}



{\bf PROGRAM SUMMARY}

\begin{small}
\noindent
{\em Program Title:FaSE}                                          \\
{\em Developer's respository link:} https://github.com/tcwphy/FASE\_GLoBES\\
{\em Licensing provisions(please choose one):} GNU General Public License 3 (GPL)\\
{\em Programming language:}C/C++                                   \\
{\em Nature of problem(approx. 50-250 words):}\\
The FaSE package serves to provide a toolkit for GLoBES to test general flavor symmetry models in neutrino oscillation experiments.
{\em Solution method(approx. 50-250 words):}
Two files are provided in FaSE: `FaSE\_GLoBES.c'  and `model-input.c'. The function of `FaSE\_GLoBES.c' is to simulate the probability profile and implement the prior values for standard neutrino oscillation parameters, which are translated from the model parameters given by the user. The user-defined input, which includes the model set up and any restrictions on the model parameters, are defined in `model-input.c'.
\\
{\em Additional comments including restrictions and unusual features (approx. 50-250 words):}
This toolkit is not a standalone program, but it requires an installation of GLoBES version 3.0.0 or higher. \\
   \\

\section{Introduction}\label{sec:intro}

The discovery of neutrino oscillations points out the fact that neutrinos have mass, and provides an evidence beyond the Standard Model (BSM). This phenomenon is successfully described by a theoretical framework with the help of three neutrino mixing angles ($\theta_{12}$, $\theta_{13}$, $\theta_{23}$), two mass-square splittings ($\Delta m_{21}^2$, $\Delta m_{31}^2$), and one Dirac CP phase ($\delta$) \cite{Pontecorvo:1967fh,Maki:1962mu,Pontecorvo:1957qd,Esteban:2018azc}. Thanks to great efforts in the past two decades, we almost have a complete understanding of such a neutrino oscillation framework. Nevertheless, more efforts in the neutrino oscillation experiments are needed to determine the sign of $\Delta m_{31}^2$, to measure the value of $\sin\theta_{23}$ more precisely, to discover the potential CP violation in the leptonic sector and even to constrain the size of $\delta$ \cite{Esteban:2018azc}. For these purposes, the on-going long baseline experiments (LBLs), such as the NuMI Off-axis $\nu_e$ Appearance experiment (NO$\nu$A)~\cite{Ayres:2007tu} and the Tokai-to-Kamioka experiment (T2K)~\cite{Abe:2011ks}, can answer these questions with the statistical significance $\gtrsim 3\sigma$ in most of the parameter space. Based on the analysis in T2K and NO$\nu$A, the normal mass ordering ($\Delta m_{31}^2>0$), the higher $\theta_{23}$ octant ($\theta_{23}>45^\circ$), and $\delta\sim270^\circ$ are preferred so far~\cite{Esteban:2018azc}. The future LBLs, Deep Underground Neutrino Experiment (DUNE)~\cite{Acciarri:2015uup}, Tokai to Hyper-Kamiokande (T2HK)~\cite{Abe:2014oxa}, and the medium baseline reactor experiment, the Jiangmen Underground Neutrino Observatory (JUNO)~\cite{Djurcic:2015vqa,An:2015jdp} will further complete our knowledge of neutrino oscillations.

The generation mechanism of neutrino masses is still a mystery in particle physics.  
Though the latest cosmological result shows the smallness of neutrino mass $\sum m_\nu<\mathcal{O}(0.1)$~eV~\cite{Aghanim:2018eyx,Giusarma:2016phn,Vagnozzi:2017ovm}, the mass of each neutrino is not clear.
In addition, the true theory that explains the origin of neutrino mass is waiting to be found out.
Models based on the seesaw mechanism have been used to explain such a tiny mass in the neutrino sector.
Furthermore, flavour symmetries can be employed to reduce degrees of freedom in the neutrino mass model. These models can explain the origin of the neutrino mixing, and predict correlations of oscillation parameters (some of recent review articles are~\cite{Altarelli:2010gt,Ishimori:2010au,King:2013eh,King:2014nza,King:2015aea,King:2015ata,King:2017guk}). 
Several neutrino mixing patterns have been proposed, such as tribimaximal mixings(TB)~\cite{Harrison:2002er,Xing:2002sw}, democratic mixings~\cite{Fritzsch:1995dj}, bimaximal mixings(BM)~\cite{Fukugita:1998vn,Barger:1998ta,Davidson:1998bi}, golden ratio mixings(GR)~\cite{Datta:2003qg,Kajiyama:2007gx,Everett:2008et,Feruglio:2011qq}, and hexogonal mixings~\cite{Albright:2010ap}. After the measurement of non-zero $\theta_{13}$, which almost excluded TB, BM, GR mixings, the surviving extension of TB mixing is mainly discussed~\cite{Harrison:2002er,King:2007pr,Pakvasa:2007zj}.
While the high energy symmetry $G_f$ is slightly broken in the lower energy, the mixing pattern is realised and the size of CP phase $\delta$ is predicted. 
As some well-known models with $A_4$ and $S_4$ symmetries can realise the TB mixing, $G_f$ contains at least one of these symmetries. There are several approaches for the symmetry breaking from the high to low energy: direct~(\textit{e.g.}~Ref.~\cite{King:2013eh}), indirect~(\textit{e.g.}~Ref.~\cite{King:2013eh}), semi-direct(\textit{e.g.}~Ref.~\cite{Holthausen:2012dk,Feruglio:2012cw,Feruglio:2013hia,Ding:2013hpa,Ding:2013bpa}), tri-direct~(\textit{e.g.}~Ref.~\cite{Ding:2018fyz,Ding:2018tuj}). All of them can explain the current data, and predict the size of $\delta$, which will be measured at the high precision in the upcoming neutrino experiments. Moreover, there are also different theories explaining the origin of the symmetry $G_f$ as well: continuous non-Abelian gauge theories \textit{such as} $SO(3)$ (\textit{e.g.}~Ref.~\cite{King:2005bj}) or $SU(3)$ (\textit{e.g.}~Ref.~\cite{King:2001uz,King:2003rf}), and the discrete symmetry from extra dimensions (\textit{e.g.}~Ref.~\cite{Feruglio:2017spp,Criado:2018thu,Penedo:2018nmg}).
Though we do not discuss each of these models in detail, we still recommend the users, who are not familiar with these models, to visit these references.

\begin{table}[h!]
\caption{\label{tab:exp}Summary of available AEDL files for some of most-discussed experiments.}
\centering
\begin{tabular}{l|c|r}
exp.   & source & ref. \\\hline\hline
T2K    &     T2K.glb on \cite{GLoBES}  &    \cite{Huber:2002mx}  \\
NOvA   &     NOvA.glb on \cite{GLoBES}   &   \cite{Ambats:2004js}   \\
T2HK   & sys-T2HK.glb on \cite{GLoBES}  &   \cite{Huber:2007em}   \\
DUNE   &    \cite{Alion:2016uaj}    &  \cite{Alion:2016uaj}    \\
MOMENT &   on FaSE website    &     \cite{Cao:2014bea}
\end{tabular}
\end{table}

It is relatively easy for model builders to check the validity of the neutrino mass model and constrain model parameters by the public NuFit results \cite{Esteban:2018azc}. However, there is no such a public toolkit to evaluate model predictions in future neutrino experiments. \textbf{G}eneral \textbf{Lo}ng \textbf{B}aseline \textbf{E}xperiment \textbf{S}imulator (\textbf{GLoBES}) \cite{Huber:2004ka,Huber:2007ji} is a convenient tool to simulate neutrino oscillation experiments via the Abstract Experiment Definition Language (AEDL). It is taken as one of the most popular and powerful simulation tools in the community of neutrino oscillation physics. Some AEDL files to describe experiments are also available in \textbf{GLoBES} website~\cite{GLoBES}, while the working group in the DUNE experiment also releases their neutrino flux information and detector descriptions in a AEDL file, provided in~\cite{Alion:2016uaj}. We summarise AEDL files for some of the most interesting experiments in Table.~\ref{tab:exp}, including their sources and references. It is to be extended for the purpose of analysing flavour symmetry models in an universal way.

As we are entering the era of precision measurements in the neutrino oscillations, recent works pay more attentions to how the future neutrino experiments can be used to test these flavour-symmetry neutrino mass models, \textit{e.g.}~Ref.~\cite{Ballett:2016yod,Agarwalla:2017wct,Chatterjee:2017xkb,Petcov:2018snn, Ding:2019zhn, Tang:2019edw, Blennow:2020snb,Blennow:2020ncm}. In this work, we will present our simulation toolkit \textbf{F}l\textbf{a}vour \textbf{S}ymmetry \textbf{E}mbedded - \textbf{GLoBES} (\textbf{FaSE-GloBES}) in a C-library to facilitate the study in the flavour symmetry neutrino models~\cite{GLoBES}. \textbf{FaSE} is a supplemental tool for \textbf{GLoBES}, written in \texttt{c/c++} language, and allows users to assign any flavour symmetry model and analyze how a flavour symmetry model is constrained by the simulated neutrino oscillation experiments.

\section{Overview FASE-GloBES}\label{sec:overview}

\textbf{FaSE} is written in \texttt{c/c++}, and consists of two source codes \textbf{FaSE\_GLoBES.c} and \textbf{model-input.c}, and is available on \url{https://github.com/tcwphy/FaSE_GLoBES}. About these two \texttt{c}-codes, the user defines the correlations between model inputs and standard neutrino mixing parameters in \textbf{model-input.c}, while \textbf{FaSE\_GLoBES.c} is a probability engine for performing an analysis with user-specified experiments in a simulation. Note that we define the standard neutrino mixing/oscillation parameters $\vec{\theta}_{OSC}$ ($\theta_{12}$, $\theta_{13}$, $\theta_{23}$, $\Delta m_{21}^2$, $\Delta m_{31}^2$, $\delta$) to separate from model parameters $\vec{\theta}_{Model}$ hereafter. 
\begin{figure}[!h]%
\centering
\includegraphics[width=5.in]{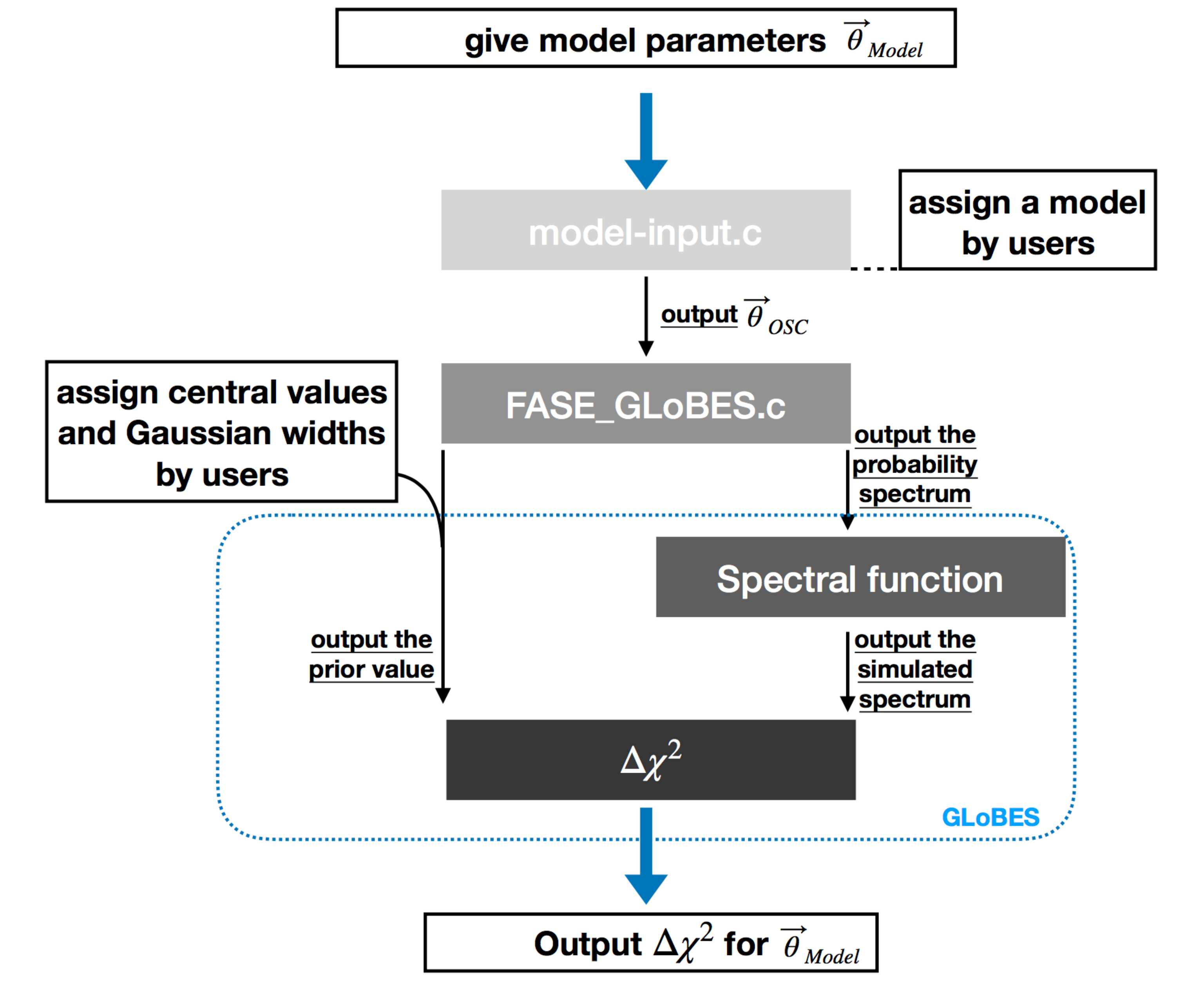}
\caption{
A sketch for the concept of \textbf{FaSE-GLoBES}. From top to down, this sketch shows three parts: 1.~\textbf{the parameter translation} (\textbf{model-input.c}), 2.~\textbf{giving oscillation-parameter values} (\textbf{FaSE\_GLoBES.c}), and 3.~\textbf{the $\chi^2$-value calculation} (\textbf{GLoBES}).
}%
\label{fig:FASE}
\end{figure}

Combining \textbf{GLoBES} with \textbf{FaSE} (we call it `\textbf{FaSE-GLoBES}'),  the user can analyse flavour symmetry models with the simulated experimental configurations. The concept of \textbf{FaSE-GLoBES} is shown in Fig.~\ref{fig:FASE}, in which three parts are shown: 1.~\textbf{the parameter translation}, 2.~\textbf{giving oscillation-parameter values}, and 3.~\textbf{the $\chi^2$-value calculation}. 
The idea behind this flow chart in Fig.~\ref{fig:FASE} is that given a set of model parameters $\vec{\theta}^{hyp}_{Model}$ as a hypothesis, the corresponding values in standard oscillation parameters $\vec{\theta}_{OSC}$ are obtained by a translation function, which is assigned by the user in \textbf{model-input.c}. And then, through \textbf{FaSE\_GLoBES.c} these oscillation-parameter values are passed into \textbf{GLoBES} library to simulate the event spectra so that the user can perform the physics analysis with the newly-defined $\chi^2$.

Application Programming Interface (API) functions in \textbf{FaSE} are listed:
\begin{enumerate}
\item \texttt{MODEL\_init($N_{para}$)},
\item  \texttt{FASE\_glb\_probability\_matrix},
\item  \texttt{FASE\_glb\_set\_oscillation\_parameters},
\item  \texttt{FASE\_glb\_get\_oscillation\_parameters},
\item \texttt{FASE\_prior\_OSC},
\item \texttt{FASE\_prior\_model}.
\end{enumerate}
The first one is to initialise \textbf{FaSE} with the number of input parameters $N_{para}$, which should not be larger than $6$. The next three functions need to be included to replace the default \textbf{GLoBES} probability engine. In the main code, the user needs to include the script as follows.
\begin{verbatim}
 glbRegisterProbabilityEngine(6,
                                 &FASE_glb_probability_matrix,
                                 &FASE_glb_set_oscillation_parameters,
                                 &FASE_glb_get_oscillation_parameters,
                                 NULL);
\end{verbatim}

This probability engine can work with oscillation or model parameters. It is set by the user with the parameter \texttt{PARA} in the main code. If \texttt{PARA=STAN} (\texttt{PARA=MODEL}) the probability engine works with oscillation (model) parameters. The final two items on the API list are prior functions. Once the user gives the prior in oscillation (model) parameters, the user needs to call \texttt{FASE\_prior\_OSC} (\texttt{FASE\_prior\_model}) as follows.
\begin{verbatim}
 glbRegisterPriorFunction(FASE_prior_OSC,NULL,NULL,NULL);
\end{verbatim}
or
\begin{verbatim}
 glbRegisterPriorFunction(FASE_prior_model,NULL,NULL,NULL);
\end{verbatim}

We note that except for setting the probability engine and the prior function, the other parts in the main code should follow with the \textbf{GLoBES} manual.

\section{Model setting}\label{sec:model_set}

The function \texttt{MtoS} can do the translation from model parameters $\vec{\theta}_{Model}$ to oscillation parameters $\vec{\theta}_{OSC}$. After the user gives the array $\vec{\theta}_{Model}$ to the function \texttt{MtoS}, the output is the corresponding oscillation parameter $\vec{\theta}_{OSC}$, of which components are $\theta_{12}$, $\theta_{13}$, $\theta_{23}$, $\delta$, $\Delta m_{21}^2$, and $\Delta m_{31}^2$. For the first four components, values are given in the unit of \textbf{rad}, while the other two are in \textbf{eV$^2$}. These values will be passed into \textbf{FaSE\_GLoBES.c} to simulate experimental spectra and compute prior values.

There are three methods to translate from $\vec{\theta}_{Model}$ to $\vec{\theta}_{OSC}$ in \textbf{FaSE-GLoBES} as follows. 

\begin{enumerate}
 \item {Assign the relation between the standard oscillation and model parameter sets by equations. In this way, the user needs to provide
 \begin{equation}
\vec{\theta}_{Model}=\vec{f}(\vec{\theta}_{OSC}), 
\end{equation}
in the function \texttt{MtoS}.
}
 \item {Give the mixing matrix in model parameter $U$. When the user gives the mixing matrix $U$ in model parameters, the corresponding mixing angles can be obtained through relations,
\begin{eqnarray}\label{eq:U_angle}
\tan\theta_{12}=\left|\frac{U_{e2}}{U_{e1}}\right|,&  \sin\theta_{13}=|U_{e3}|,& \tan\theta_{23} = \left|\frac{U_{\mu 3}}{U_{\tau 3}}\right|.
\end{eqnarray}
After getting all mixing angles, we can easily derive the Dirac CP phase $\delta$ with the Jarlskog invariant $J_{CP}$,\
\begin{equation}\label{eq:U_phase}
J_{CP}=c_{12}s_{12}c_{23}s_{23}c_{13}^2s_{13}\sin\delta,
\end{equation}
where $c_{ij}$ and $s_{ij}$ are $\cos \theta_{ij}$ and $\sin\theta_{ij}$, respectively. In \texttt{MtoS}, the user can pass the mixing matrix $U$ to the function  \texttt{STAN\_OSC\_U} to obtain corresponding mixing angles and the CP phase. 
}
 \item {Define the mass matrix in model parameters alternatively. The oscillation parameters can also be obtained in the way based on
\begin{equation}\label{eq:MM}
U^\dagger\mathcal{M}\mathcal{M}^\dagger U = \mathbf{M}^2,~\text{where}~\mathbf{M}^2_{\alpha\beta}=m_\alpha^2\delta_{\alpha\beta},
\end{equation}
where $\mathcal{M}$ ($\mathbf{M}$) is the neutrino mass matrix in the flavour (mass) state. The matrix $\mathcal{M}$ is given by the user with model parameters $\vec{\theta}_{Model}$. The mixing matrix $U$ can be used to get mixing angles and the CP phase, as Eqs. (\ref{eq:U_angle}) and (\ref{eq:U_phase}). The difference between any two diagonal elements of $\textbf{M}$ ($\textbf{M}_{ii}-\textbf{M}_{jj}=m_i^2-m_j^2$) is the mass-squared difference ($\Delta m_{ij}^2$). This diagnolisation in Eq.~(\ref{eq:MM}) can be done by the function \texttt{STAN\_OSC}, which needs to be called in \texttt{MtoS} with outputs of the vector $\vec{\theta}_{OSC}$. We note that with non-diagonal mass matrix for charged leptons in models, this translation method is not suggested.}
\end{enumerate}

\section{Prior setting}\label{sec:prior}
Given a set of values for model parameters, \textbf{FaSE\_GLoBES.c} will obtain the corresponding oscillation-parameter values from \textbf{model-input.c}, and will pass these values to simulate event spectra and to compute the prior values. Two gaussian prior functions are provided in \textbf{FaSE}: \texttt{FASE\_prior\_OSC} and \texttt{FASE\_prior\_model}. These two functions are constructed for different purposes. If the user gives the prior in oscillation (model) parameters, the user should register \texttt{FASE\_prior\_OSC} (\texttt{FASE\_prior\_model}) with the \textbf{GLoBES} function \texttt{glbRegisterPriorFunction}, as we introduced in Sec.~\ref{sec:overview}. The user also needs to assign the parameters \texttt{PARA=STAN} (\texttt{PARA=Model}), when the user prefers to give the prior in oscillation (model) parameters. The Gaussian prior is 
\begin{equation}\label{eq:prior}
\chi^2_{prior}=\sum_{i} \frac{(\theta_i-\theta^c_i)^2}{\sigma_i^2},
\end{equation}
 where $\theta_i$ is one parameter of the hypothesis $\vec{\theta}^{hyp}$, $\theta^c_i$ ($\sigma_i$) is the central value (Gaussian width) of the prior for $\theta_i$. We note that $\vec{\theta}^{hyp}$ can be either model ($\vec{\theta}_{Model}$) or oscillation parameters ($\vec{\theta}_{OSC}$).
The values of $\theta^c_i$ and $\sigma_i$ need to be given by the user in the main code through three arrays: \texttt{Central\_prior}, \texttt{UPPER\_prior}, and \texttt{LOWER\_prior}, in which there are six components. To treat asymmetry of width for upper ($\theta_i>\theta_i^c$) and lower ($\theta_i<\theta_i^c$) Gaussian widths, we give values in two arrays \texttt{UPPER\_prior}, and \texttt{LOWER\_prior}, respectively. Once the user does not want to include any priors, two arrays \texttt{UPPER\_prior} and \texttt{LOWER\_prior} need to be $0$. If the user gives the prior in model parameters, the order of each component follows with the setup of input of the probability engine. While the user gives the prior in oscillation parameters, the six components of these three arrays in order are $\theta_{12}$, $\theta_{13}$, $\theta_{23}$, $\delta$, $\Delta m_{21}^2$, and  $\Delta m_{31}^2$. The first four parameter are in \textbf{rad}, and the final two are in \textbf{eV$^2$}.

Finally, some restrictions are imposed by the chosen flavour symmetry model. We set up these restrictions in the function \texttt{model\_restriction} in \texttt{model-input.c}. In the function \texttt{model\_restriction}, the user needs to \textit{return $1$} once the restriction is broken. For example, if the normal ordering is imposed, we give ``\texttt{if (DMS31<0) \{ return 1;\} }'' in \texttt{model\_restriction}, where \texttt{DMS31} is the variable for $\Delta m_{31}^2$. 
And, if there is no restriction, we simply \textit{return $0$} in \texttt{model\_restriction} as follows:
\begin{verbatim}
 double model_restriction(double model []){ return 0;}.
\end{verbatim}


\section{The definition for $\chi^2$ (based on \textbf{GLoBES})}\label{sec:chi-squared}

The user can use \textbf{FaSE-GLoBES} to constrain model parameters. 
Suppose we have the measurement $\vec{x}^{hyp}$ and the likelihood function $L(\vec{\theta}^{hyp})=P(\vec{x}|\vec{\theta}^{hyp})$ for a set of parameters $\vec{\theta}^{hyp}=(\theta_1,...,\theta_N)$, where $P(\vec{x}|\vec{\theta}^{hyp})$ is the probability function for data $\vec{x}$ in favour of the hypothesis $\vec{\theta}^{hyp}$. The constraint of model parameters can be obtained with the statistis parameter $\chi^2\equiv \ln L(\vec{\theta}^{hyp})$.
The expression $\chi^2$ is used as the default \textbf{GLoBES} setting. In more detail, the $\chi^2$ function, following the Poisson distribution, is constructed based on a log-likelihood ratio,
\begin{align}\label{eq:chi-squared}
\chi^2(\vec{\theta}^{hyp},\xi_s,\xi_b)=&2\sum_i\left(\eta_i(\vec{\theta}^{hyp},\xi_s,\xi_b)-n_i+n_i\ln\frac{n_i}{\eta_i(\vec{\theta}^{hyp},\xi_s,\xi_b)} \right)\nonumber\\
&+p(\xi_s,\sigma_s)+p(\xi_b,\sigma_b)+\chi^2_{prior},
\end{align}
where $i$ runs over the number of bins, $\eta_i(\vec{\theta},\xi_s,\xi_b)$ is the assumed event rate in the $i$th bin and $E_i$ is the central value in this energy bin. The vector $\vec{\theta}$ consists of model or oscillation parameters. The parameters $\xi_s$ and $\xi_b$ are introduced to account for the systematic uncertainties in the normalisation for the signal (subscript $_s$) and background (subscript $_b$) components of the event rate, and are allowed to vary in the fit as nuisance parameters. For a given set of parameters $\vec{\theta}^{hyp}$, the event rate in the $i$th energy bin is calculated as\\
\begin{equation}
\eta_i(\vec{\theta}^{hyp},\xi_s,\xi_b)=(1+\xi_s)\times s_i+(1+\xi_b)\times b_i,
\end{equation}
where $s_i$ and $b_i$ are the expected number of signal and background events in $i$th energy bin, respectively. The nuisance parameters are constrained by the Gaussian prior $p(\xi,\sigma)=\xi^2/\sigma^2$ with corresponding uncertainties $\sigma_s$ and $\sigma_b$ for the signal and background, respectively. Finally, $\chi^2_{prior}$ is a set of Gaussian priors for hypothesis, and is expressed as Eq.~(\ref{eq:prior}). After doing all minimisations, the user obtains the $\chi^2$ value for a specific hypothesis $\vec{\theta}^{hyp}$, $\chi^2(\vec{\theta}^{hyp})$.

Based on the $\chi^2$ function Eq.~(\ref{eq:chi-squared}), we can study how model parameters can be constrained and whether a flavour-symmetry neutrino model is excluded by simulated experiments. In the following we will demonstrate with typical examples how it works, before presenting some demonstrations in next two sections.

\subsection*{Applications}

The user of \textbf{FaSE-GLoBES} is able to study how model parameters can be constrained by the simulated experiments. To do so, the user needs to simulate the true event spectrum $n_i$ with a set of model ($\vec{\theta}_{Model}^{true}$) or oscillation parameters ($\vec{\theta}_{OSC}^{true}$), \textit{i.e.}~set up $n_i(\vec{\theta}_{Model}^{true})$ or $n_i(\vec{\theta}_{OSC}^{true})$. The hypothesis $\vec{\theta}_{Model}^{hyp}$ predicts the tested event spectrum $\eta_i(\vec{\theta}_{Model}^{hyp},\xi_s,\xi_b)$. With the default settings for $\chi^2$ function as Eq.~(\ref{eq:prior}) in \textbf{FaSE-GLoBES}, the user computes the statistical quantity,
\begin{equation}\label{eq:chi_model}
\chi^2(\vec{\theta}_{Model}^{hyp}),~~\text{with}~n_i(\vec{\theta}_{Model}^{true})~\text{or}~n_i(\vec{\theta}_{OSC}^{true}).
\end{equation}
We note that the minimum of $\chi^2$ in the whole parameter space ($\chi^2_{min.}$) may not be $0$. Therefore, to get the precision of model parameters, the user should use the value $\Delta\chi^2(\vec{\theta}_{Model}^{hyp})\equiv \chi^2(\vec{\theta}_{Model}^{hyp})-\chi^2_{min.}$, instead of $\chi^2(\vec{\theta}_{Model}^{hyp})$ itself. By varying different hypotheses $\vec{\theta}_{Model}^{hyp}$, we will obtain the allowed region of model parameters with the statistical quantity $\Delta\chi^2(\vec{\theta}_{Model}^{hyp})$. 

The user can also study how well a flavour symmetry model explains the computed data, or predict whether the simulated experiment can exclude this model or not. In other words, the user studies the minimum of $\chi^2$ for the flavour symmetry model $\vec{\theta}_{Model}$ as a hypothesis, by assuming different true oscillation values, \textit{i.e.}~different $\vec{\theta}^{true}_{OSC}$. To do so, one can compute the same statistical quantity in Eq.~(\ref{eq:chi_model}), while the true spectrum is varied with different true values $\vec{\theta}_{OSC}^{true}$. All model parameters are allowed to be varied with the user-defined prior.
Finally, the user might adopt Wilk's theorem to interpret results~\cite{Wilks:1938dza}. When we compare nested models, the $\Delta \chi^2$ test statistics is a random variable asymptotically distributed as a $\chi^2$-distribution with the number of degrees of freedom, which is equal to the difference in the number of free model parameters. 

In following two sections, we will present examples to demonstrate how the user can make use of \textbf{FaSE-GLoBES} to constrain the model parameter and to exclude a model by the simulated experiment configurations.

\section{Constraint of model parameters}

\textbf{FaSE-GloBES} can be used to study how model parameters are constrained by simulated neutrino oscillation experiments as we introduced in Sec.~\ref{sec:chi-squared}. We take \textit{the tri-direct littlest seesaw} (TDLS)~\cite{King:2013iva,King:2015dvf,King:2016yvg} as an example. In this model, the light left-handed Majorana neutrino mass matrix in the flavor basis is given by
\begin{equation}
\label{eq:mnu}  m_{\nu}=m_{a}\begin{pmatrix}
 1 &~ \omega  &~ \omega ^2 \\
 \omega  &~ \omega ^2 &~ 1 \\
 \omega ^2 &~ 1 &~ \omega  \\
\end{pmatrix}+e^{i\eta}m_{s}
\begin{pmatrix}
 1 &~  x &~  x \\
 x &~ x^2 &~ x^2 \\
 x &~ x^2 &~ x^2 \\
\end{pmatrix}\,,
\end{equation}
where $x$, $\eta$, $m_a$, and the ratio $r\equiv m_s/m_a$ are four parameters to be constrained by simulated data. We note that from Eq.~(\ref{eq:mnu}), $m_1=0$ and the normal mass ordering are imposed, and will need to be imposed in \textbf{FaSE-GLoBES}. Therefore, the restrictions in this model are $m_a>0$ and $r>0$.

\begin{figure}[!h]
 \centering
\includegraphics[width=0.48\textwidth]{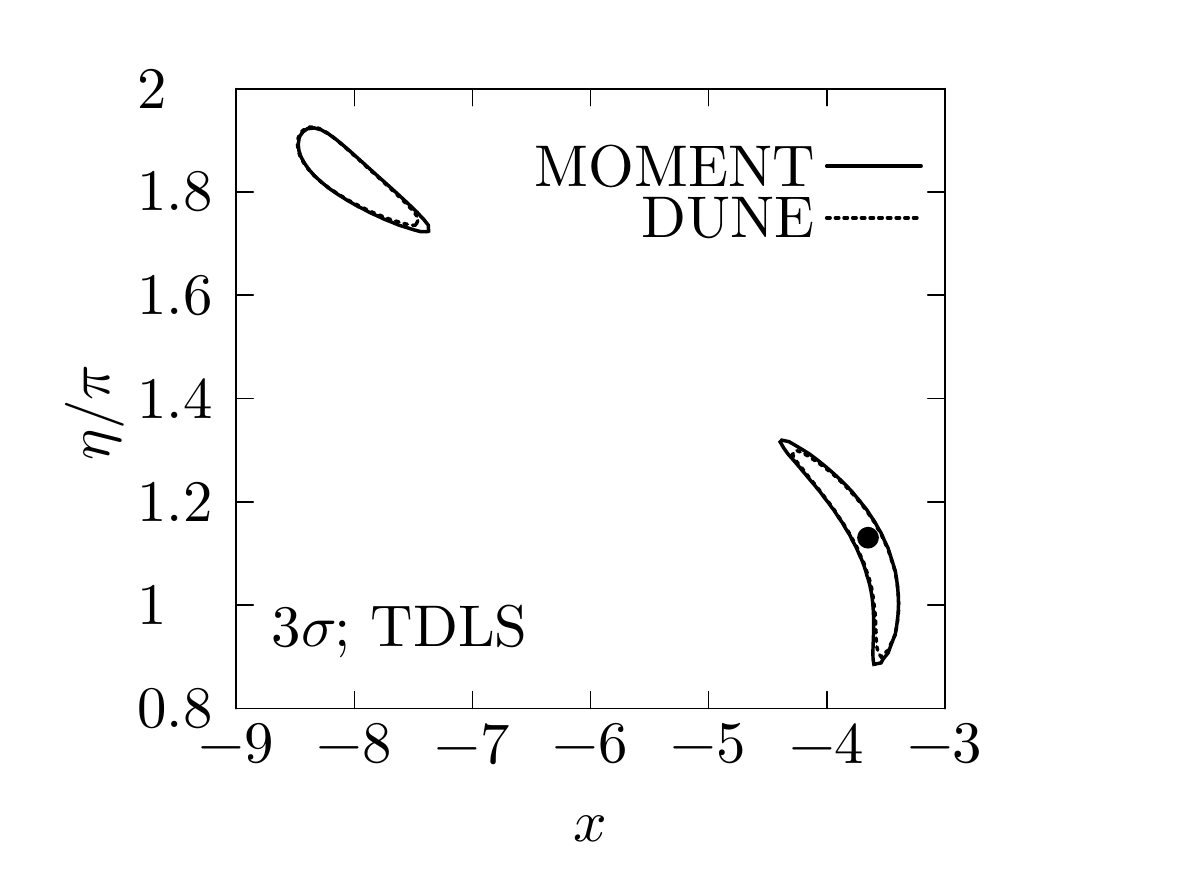}
 \includegraphics[width=0.48\textwidth]{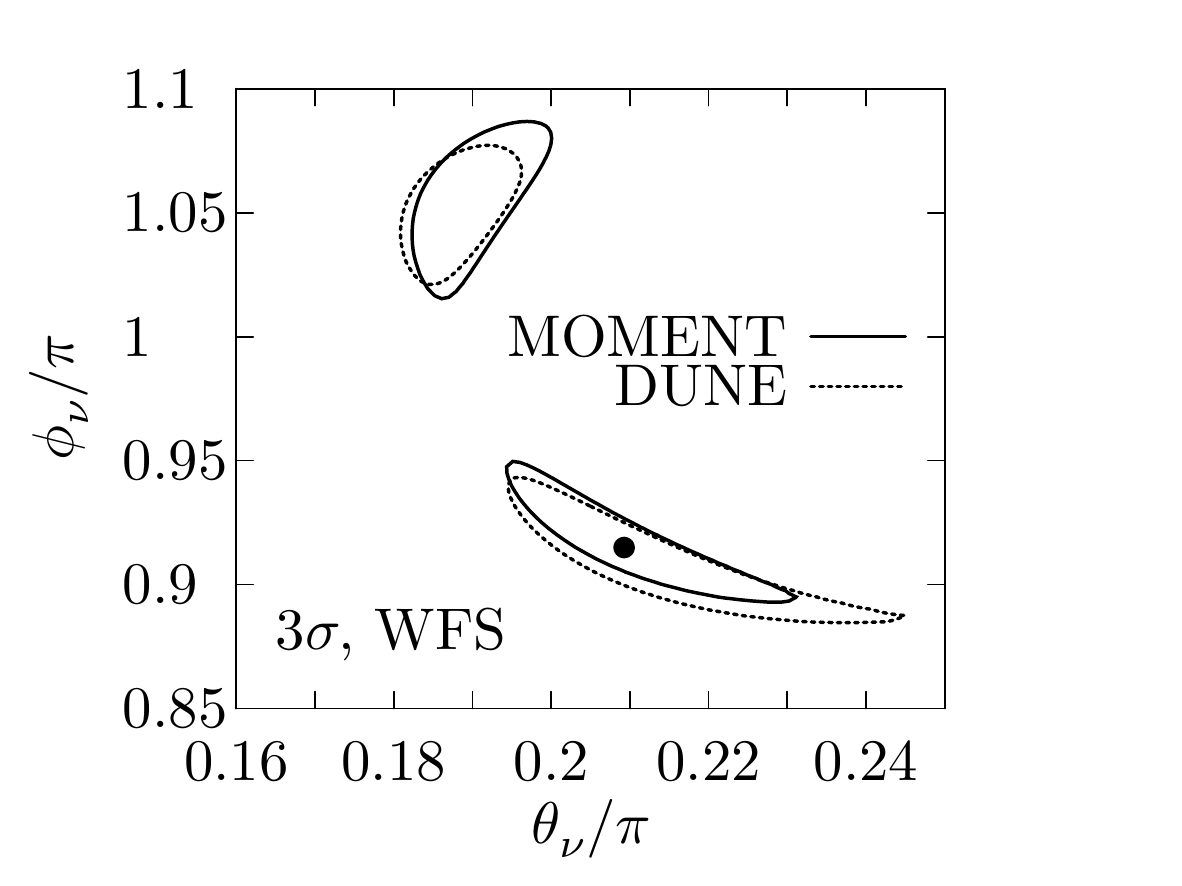}
 %
  \caption{\label{fig:constraint}Examples for using \textbf{FaSE-GLoBES} to obtain the constraints of model parameters for \textit{tri-direct littlest seesaw} (left) and \textit{the warped flavor symmetry} (right), with simulated DUNE and MOMENT data. Two results are assumed the normal ordering. The black dot denotes the model prediction with NuFit4.0 results.}
\end{figure}

In the left panel of Fig.~\ref{fig:constraint}, we study how model parameters $x$ and $\eta$ can be constrained at $3\sigma$ C.L. by the MuOn-decay MEdium baseline NeuTrino beam experiment (MOMENT)~\cite{Cao:2014bea} and DUNE experiment. 
Parameters $r$ and $m_a$ are varied with the prior that is given in standard oscillation parameters, according to the global-fit result NuFit4.0. 


To show the generality of \textbf{FaSE-GLoBES}, we also present the similar result for another model -- \textit{the warped flavor symmetry} (WFS)~\cite{Chen:2015jta}. This model predicts further simplified correlations that the standard oscillation parameters including mixing angles and the CP phase are functions of only two model parameters $\theta_\nu$ and $\phi_\nu$,
\begin{align}\label{eq:WFS}
\sin^2\theta_{12}&=\frac{1}{2-\sin2\theta_{\nu}\cos\phi_\nu},\nonumber\\
\sin^2\theta_{13}&=\frac{1}{3}(1+\sin2\theta_\nu\cos\phi_\nu),\nonumber\\
\sin^2\theta_{23}&=\frac{1-\sin2\theta_\nu\sin(\pi/6-\phi_\nu)}{2-\sin2\theta_\nu\cos\phi_\nu},\nonumber\\
J_{CP}&=-\frac{1}{6\sqrt{3}}\cos2\theta_\nu.
\end{align}
The constraint of $\theta_\nu$ and $\phi_\nu$ for DUNE and MOMENT is presented in the right panel of Fig~\ref{fig:constraint}, in which we use the best fit of NuFit 4.0 result as the true values $(\theta_{12},~\theta_{13},~\theta_{23},~\delta,~\Delta m_{21}^2,~\Delta m_{31}^2)=(33.82^\circ,~8.61^\circ,~49.6^\circ,~215^\circ,~7.39\times10^{-5}~\text{eV}^2,~2.525\times10^{-3}~\text{eV}^2)$. To reproduce results shown in \cite{Chatterjee:2017xkb}, we do not include any priors. More details about these codes are presented in the user manual\footnote{The manual is available on the \textbf{FaSE} repository  \url{https://github.com/tcwphy/FaSE_GLoBES/doc}.}.

\section{Model testing}
\begin{figure}[!h]
 \centering
\includegraphics[width=0.48\textwidth]{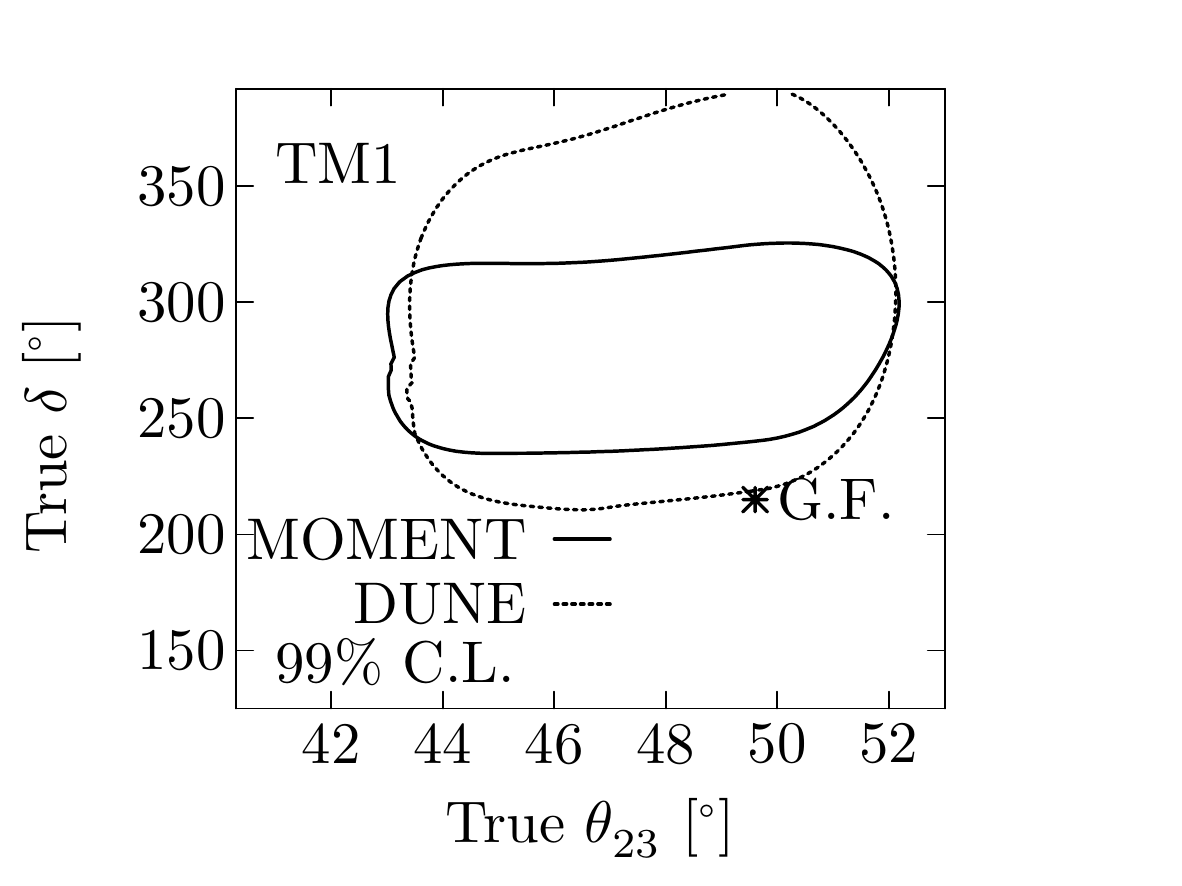}
\includegraphics[width=0.48\textwidth]{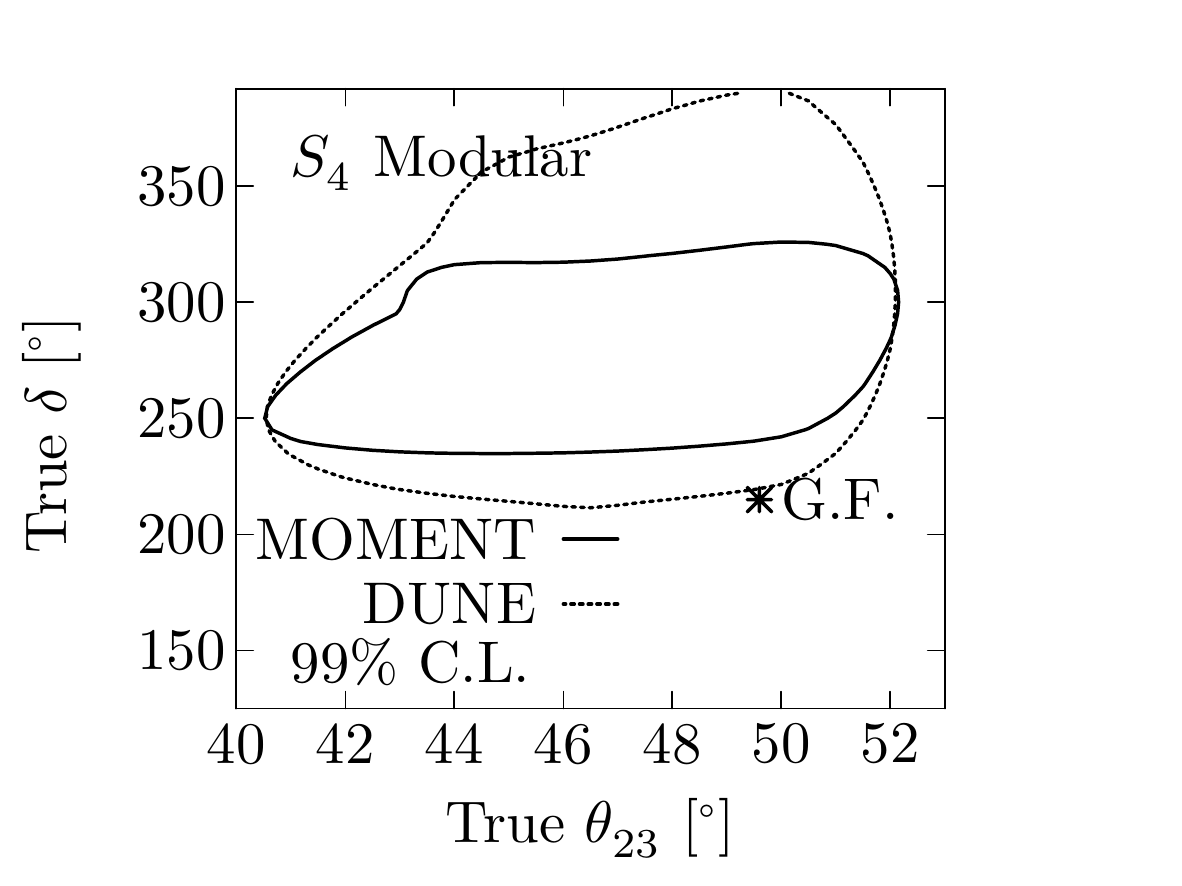}
\caption{\label{fig:th23_delta}Examples for using \textbf{FaSE-GLoBES} to obtain the 2-D exclusion contour at $99\%$ C.L. on the plane of true values of $\theta_{23}$ and $\delta$ for \textit{TM1} sum rule (left) and \textit{a $S_4$ modular symmetry model} (right) with DUNE and MOMENT. }
\end{figure}

We can also study on how much a neutrino mass model  or a sum rule can be excluded, assuming different true values for oscillation parameters.
In Fig.~\ref{fig:th23_delta}, we present testing \textit{trimaximal mixing TM1}~\cite{Albright:2008rp,Xing:2006ms} (left) and \textit{a $S_4$ modular symmetry model}~\cite{deMedeirosVarzielas:2019cyj} (right) in various true values of $\theta_{23}$ and $\delta$. TM1 implies three equivalent relations between $\theta_{12}$ and $\theta_{13}$~\cite{Albright:2008rp} :
\begin{eqnarray}\label{eq:TM1}
\tan\theta_{12}=\frac{1}{2}\sqrt{1-3s^2_{13}},&\sin\theta_{12}=\frac{1}{3}\frac{\sqrt{1-3s_{13}^2}}{c_13},&\text{or}~\cos\theta_{12}=\sqrt{\frac{2}{3}}\frac{1}{c_{13}}.
\end{eqnarray}
and also the dependence of $\delta$ on $\theta_{13}$ and $\theta_{23}$:
\begin{eqnarray}\label{eq:TM1_dCP}
\cos\delta=-\frac{\cot 2\theta_{23}(1-5s_{13}^2)}{2\sqrt{2}s_{13}\sqrt{1-3s_{13}^2}}.
\end{eqnarray}
The other model, we use for demonstration, is based on three moduli with finite modular symmetries $S_4^A$, $S_4^B$, and $S_4^C$, associated with two right-handed neutrinos and the charged lepton sector, respectively~\cite{deMedeirosVarzielas:2019cyj}. This model predicts the neutrino mass matrix: 
\begin{eqnarray}
m_\nu=&(\mu_1\hat{c}_R^2+\mu_2\hat{s}_R^{*2})\left(
\begin{array}{ccc}
1& -2\omega^2& -2\omega\\
-2\omega^2&4\omega&4\\
-2\omega&4&4\omega^2
\end{array}
\right)+
(\mu_1\hat{s}^2_R+\mu_2\hat{c}_R^{*2})\left(
\begin{array}{ccc}
0& 0& 0\\
0&1&-1\\
0&-1&1
\end{array}
\right) \nonumber \\
&+(\mu_1\hat{c}_R\hat{s}_R-\mu_2\hat{c}^*_R\hat{s}_R^*)\left(
\begin{array}{ccc}
0& -1& 1\\
-1&4\omega^2&2i\sqrt{3}\\
1&2i\sqrt{3}&-4\omega
\end{array}
\right),
\end{eqnarray}
where $\hat{c}_R$ and $\hat{s}_R$ are $\cos\theta_R\times \mathrm{e}^{i\alpha_2}$ and $\sin\theta_R\times \mathrm{e}^{i\alpha_3}$, respectively. As $\omega$ is fixed at $-\frac{1}{2}+i\frac{\sqrt{3}}{2}$, this model has 5 model parameters: $\mu_1,~\mu_2,~\theta_R,~\alpha_2,~\alpha_3$.

We compute the minimal $\chi^2$ value for the model, and allowed all model parameters varied with the priors defined in Eq.~(\ref{eq:prior}) according to NuFit4.0 results.
In addition, the studied statistics function is exactly given by Eq.~(\ref{eq:chi-squared}), the true event rate $n_i$ is predicted by a set of assumed oscillation parameters. Two parameters $\theta_{23}$ and $\delta$ in $\vec{\theta}^{true}_{OSC}$ keep varied in the range of $40^\circ<\theta_{23}<53^\circ$ and $125^\circ<\delta<390^\circ$, respectively.
More details are presented in the user manual\footnote{The manual is available on the \textbf{FaSE} repository  \url{https://github.com/tcwphy/FaSE_GLoBES/doc}.}.

\section{Summary and conclusions}

With the progress of precision measurements in the neutrino experiments, and the success of numerous flavour symmetry theories to explain tiny neutrino masses, there are strong motivations to test and discriminate theoretical models by the next-generation neutrino oscillation experiments. We have presented a simulation toolkit \textbf{FaSE-GLoBES} to study the leptonic flavour symmetry models with neutrino oscillation experiments in a user-friendly way. \textbf{FaSE-GLoBES} contains two \texttt{c}-codes: \textbf{model-input.c} and \textbf{FaSE\_GLoBES.c}. While \textbf{FaSE\_GLoBES.c} works as a bridge between models and standard neutrino mixings, all inputs from the user need to be given in \textbf{model-input.c}. With the help of two main functions provided by \textbf{FaSE-GLoBES}, it is convenient to assign a flavour symmetry model and include Gaussian priors associated with oscillation or model parameters. Users are able to study how a flavour model can be examined by the simulated experimental configurations in various perspectives, \textit{e.g.} model parameter constraints, hypothesis testing. \textbf{FaSE-GLoBES} will contribute to the selection and screening of underlying neutrino mass models by oscillation experiments. Further improvements and extensions can be envisioned as more requests come up in model buildings and phenomenology.


\section*{Acknowledgements}
We thank Gui-Jun Ding and Ye-Ling Zhou for helpful discussions, and also thank Sampsa Vihonen for help in the code review.
This work was supported in part by Guangdong Basic and Applied Basic Research Foundation under Grant No. 2019A1515012216, National Natural Science Foundation of China under Grant Nos. 11505301 and 11881240247, the university funding based on National SuperComputer Center-Guangzhou. Jian Tang acknowledge the support from the CAS Center for Excellence in Particle Physics (CCEPP). Tse-Chun Wang was supported in part by Postdoctoral recruitment program in Guangdong province. 

\end{small}


\bibliographystyle{elsarticle-num}
\bibliography{Fase-GLoBES.bib}







\end{document}